\documentclass[11pt]{article}
\usepackage[margin=1in]{geometry}
\usepackage{amsmath,amssymb,graphicx,url}
\usepackage{authblk}

\title{\textbf{An Implementation of the Crack Topology Score\\with Extensions}}

\author[1]{Siheon Joo}
\author[1]{Hongjo Kim\thanks{Corresponding author: hongjo@yonsei.ac.kr}}
\affil[1]{Department of Civil and Environmental Engineering, Yonsei University}

\begin{document}

\maketitle

\begin{abstract}
The Crack Topology Score (CTS) is a recently proposed metric that focuses on evaluating the topological correctness of crack segmentation outputs. While pixel-wise metrics such as IoU or F1-score fail to capture structural validity, CTS offers a skeleton-based matching fframework weighted by crack length to measure the preservation of connectivity. This paper presents a faithful implementation of the CTS metric, along with optional preprocessing extensions designed to handle common prediction artifacts (e.g., small holes and edge noise) found in deep learning outputs. All extensions are disabled by default to ensure strict comparability with the original definition. The implementation supports PyTorch-based workflows and includes visualization tools for transparency. Code and archival resources will be made available at \url{https://github.com/SH-Joo/crack-topology-score}.
\end{abstract}

\section{Introduction}
Accurate evaluation of road crack segmentation is essential for real-world deployment of deep learning models in infrastructure monitoring. Traditional pixel-level metrics (e.g., IoU, Dice, pixel accuracy) do not penalize topological errors such as false disconnections or spurious branches. To address this, the Crack Topology Score (CTS) was introduced by Hyeon \textit{et al.}~\cite{Hyeon2025CTS} to emphasize structural consistency using skeleton-based comparison.

This work provides an open-source implementation of CTS, fully consistent with the metric's original formulation. In addition, we propose optional preprocessing steps---hole filling and morphological smoothing---to suppress irrelevant structural artifacts caused by minor prediction noise. These extensions are designed to align with CTS's core philosophy: focusing on topology while tolerating minor shape inaccuracies.

\section{Overview of Crack Topology Score}
Given a ground-truth binary crack mask and a predicted mask, CTS compares their skeletons as sets of curve segments. Each skeleton is decomposed into segments defined by end-point junctions. For each predicted segment, a match is declared if it overlaps with a ground-truth segment under a buffer of radius $r$ (default 10 pixels) and an overlap threshold (default 0.5).

The metric is composed of:
\begin{itemize}
  \item PCS: length-weighted proportion of predicted segments matched
  \item RCS: length-weighted proportion of ground-truth segments recovered 
  \item CTS: harmonic mean of PCS and RCS
\end{itemize}

CTS captures both completeness and correctness of predicted structures. Further details are available in the original paper~\cite{Hyeon2025CTS}.

\section{Reference Implementation}
Our implementation follows the original formulation exactly. We use Zhang--Suen thinning for skeletonization, and 8-connectivity for defining segment endpoints. The matching process applies dilation to produce buffer zones. It also integrates multiple fragmented candidate segments into a single mask to handle one-to-many overlaps correctly, strictly following the original algorithm's logic (List A integration).

The code is modular and supports PyTorch tensors. The core function takes binary masks and returns PCS, RCS, and CTS values, along with optional visualizations. When all extensions are disabled, the results are numerically consistent with the reference CTS definition.

\section{Optional Extensions}

\subsection{Hole Filling}
Small internal holes in predicted masks can cause false disconnections in the skeleton. We use flood-fill to identify background regions that are fully enclosed by foreground pixels and fill them selectively if their area is below a threshold (default: 0 = off).

This step is optional and preserves topology in most practical cases. Large holes or boundary-connected gaps are ignored.

\subsection{Morphological Smoothing}
Edge noise in thick or irregular predictions can lead to skeleton spurs, which create false branches. To address this, we apply morphological operations (open/close) before skeletonization. This pre-skeleton smoothing suppresses artifacts without modifying the final skeleton post hoc.

Users can select `open` or `close` with a disk radius (default: 0 = off). This option is useful for enhancing evaluation stability on noisy outputs.

\section{Usage and Configuration}
The full implementation, including demo scripts and configuration examples, is available at:
\begin{center}
\url{https://github.com/SH-Joo/crack-topology-score}
\end{center}

\section{Conclusion}
We present an official implementation of the Crack Topology Score metric, consistent with its original definition. Optional extensions are provided to handle typical segmentation artifacts while preserving topological evaluation intent. The tool aims to support fair and interpretable evaluation of deep learning models for crack detection.

\bibliographystyle{plain}

\end{document}